\documentclass[a4paper]{article} 

\usepackage{amsmath,amssymb}
\usepackage{graphicx}

\begin{document}

\title{\bf Bifurcating Branches of Multiple Charged One-Plus-Half Monopole In SU(2) Yang-Mills-Higgs Theory} 

\author{{Dan Zhu, Khai-Ming Wong, Timothy Tie, Guo-Quan Wong}\\
\textit{{\small School of Physics, Universiti Sains Malaysia, 11800 USM, Penang, Malaysia}}}

\date{June 2021}
\maketitle

\begin{abstract}
In this paper, we study on the one-plus-half monopole configuration in SU(2) Yang-Mills-Higgs theory when the $\phi$-winding number, $n$, runs from 2 to 4 and for a range of Higgs coupling constant, $\lambda_b \leq \lambda \leq 40$, where $\lambda_b$ is the lower bound, below which no solution can be found. Bifurcation and transition are observed for $n > 2$ when the Higgs coupling constant is larger than some critical value $\lambda_c$ and transitional value $\lambda_t$, respectively. Two different branches with energy higher than the fundamental solution are observed for both $n = 3$ and $4$. We also observed a new branch with even higher energy for $n = 4$. Unlike other branches which display transition behavior, the new branch corresponds to a full vortex-ring configuration. All the solutions possess finite energy. Plots of magnetic charge density, Higgs modulus and energy density are presented and analyzed.
\end{abstract}

\section{Introduction}
\label{sec:1}

The SU(2) Yang-Mills-Higgs theory possesses a large variety of monopole configurations and they have been studied extensively since mid-70s~\cite{r1,r2,r3,r4,r5}. No exact monopole solution has been found when the Higgs field self-coupling constant, $\lambda$, is non-vanishing~\cite{r1}. There are many different numerical monopoles and antimonopoles solutions have been found, which include monopole-antimonopole pair (MAP), monopole-antimonopole chain (MAC), and vortex ring solutions are a few common types~\cite{r6}. Most of the  monopole solutions reported are of integer magnetic charges. However, there are solutions which correspond to magnetic monopole carrying half-integer charge (coined as half-monopoles) \cite{r7}, as well as the coexistence of a full monopole and a half-monopole (coined as one-plus-half monopole solution) \cite{r8}. These configurations nevertheless possess gauge potentials that are singular only along the $z$-axis.

Here, we further investigate the electrically neutral one-plus-half monopole solution \cite{r8} when $n = 2, 3$ and $4$ for Higgs coupling constant, $\lambda_b \leq \lambda \leq 40$, where $\lambda_b$ is the lower bound, below which no solution can be found. The two bifurcating branches that possess higher energy than the fundamental branch (FB) solution are labeled as higher energy branch (HEB) and lower energy branch (LEB)  We also observed another branch that emerged when $n = 4$ and is coined as new branch (NB) for later references. Transitions from multimonopole solutions to vortex ring are observed in the HEB and LEB solutions when $n=3$. Unlike other branches which display transition behavior, the new branch corresponds to a full vortex-ring configuration only. The values for $ \lambda_b$, $\lambda_c$ and $\lambda_t$ are tabulated. Total energy $E$ and magnetic dipole moment $\mu_m$ for different branches are plotted against $\lambda^{1/2}$. Plots of Higgs modulus, magnetic charge density and energy density are also presented.

\section{The SU(2) Yang-Millls-Higgs Theory}
\label{sec:2}

The Lagrangian in $3+1$ dimensions with non-vanishing Higgs potential is
\begin{equation}\label{eq1}
\mathcal{L} = -\frac{1}{4}F^a_{\mu\nu}F^{a\mu\nu} - \frac{1}{2} D^\mu \Phi^a D_\mu \Phi^a - \frac{1}{4}  \lambda \left[\Phi^a\Phi^a -  \frac{\mu^2}{\lambda}\right]^2
\end{equation} 
where $F^a_{\mu\nu}$ is the gauge field strength tensor, $D_\mu \Phi^a$ is the covariant derivative of the Higgs field, $\lambda$ is the Higgs potential and $\xi$ is defined as $\xi = \mu/\sqrt{\lambda}$, which is the expectation value of the Higgs field. Here $\mu$ is the Higgs field mass. The Lagrangian~\eqref{eq1} is gauge invariant and stays unchanged under independent local SU(2) transformations. Parameter $a$, $b$ and $c$, which are $SU(2)$ internal group indices, run from $1$ to $3$, whereas $\mu$ and $\nu$ are space-time indices of Minkowski space, run from $0$ to $3$.

The covariant derivative of Higgs field and gauge field strength tensor are given by
\begin{eqnarray}\label{eq2}
D_\mu\Phi^a &=& \partial_\mu\Phi^a + g\varepsilon^{abc}A^b_\mu\Phi^c, \nonumber\\ 
F^a_{\mu\nu} &=& \partial_\mu A^a_\nu - \partial_\nu A^a_\mu+g\varepsilon^{abc}A^b_\mu A^c_\nu.
\end{eqnarray}
Here $g$ is gauge field coupling constant and $A^a_\mu$ is the gauge potential. Applying the Euler-Lagrange equation to Lagrangian~\eqref{eq1}, we obtain the equations of motion
\begin{eqnarray}\label{eq3}
D^\mu F^a_{\mu\nu} &=& g\varepsilon^{abc}\Phi^b D_\nu\Phi^c, \nonumber\\
D^\mu D_\mu\Phi^a &=& \lambda\Phi^a(\Phi^b\Phi^b-\xi^2).
\end{eqnarray}

The magnetic field, which can be decomposed into gauge and Higgs parts, are given by
\begin{eqnarray}\label{eq4}
&& B_i = -\frac{1}{2}\varepsilon_{ijk}F_{jk} = B^G_i + B^H_i, \nonumber\\
&& B^G_i = -n\varepsilon_{ijk}\partial_j \sin\kappa~ \partial_j\phi,~~~ B^H_i = -n\varepsilon_{ijk}\partial_j\sin\alpha~\partial_k\phi \nonumber\\
&& \sin\kappa=\frac{\sin\theta}{n} \left[ \psi_2 \frac{\Phi_2}{|\Phi|} - R_2 \frac{\Phi_2}{|\Phi|} \right], \nonumber\\
&& \sin\alpha= \frac{\Phi_1}{|\Phi|}  \cos\theta - \frac{\Phi_2}{|\Phi|}  \sin\theta
\end{eqnarray}
and thus the net magnetic charge of the system is
\begin{eqnarray}
M=  \frac{1}{4\pi}\int\partial^i B_i d^3 x = \frac{1}{4\pi}  \oint d^2 \sigma_i B_i
\label{eq5}
\end{eqnarray}

From Maxwell electromagnetic theory, 't Hooft's gauge potential that can be determined from Eq.\eqref{eq4} at large $r$ tends to
\begin{align}\label{eq6}
A_i = (\cos\alpha+\cos\kappa)\partial_i\phi|_{r\to \infty} = \frac{\hat{\phi}_i }{r\sin\theta} \left[\frac{1}{2} (\cos\theta \pm 1) +  \frac{F_G(\theta)}{r}\right]   \nonumber\\
F_G(\theta) = r \left[ \frac{\Phi_2}{|\Phi|} (P_1-\sin\theta) - \frac{\Phi_1}{|\Phi|} (P_2-\cos\theta)-\frac{1}{2}(\cos\theta\pm 1)  \right] |_{r\to\infty}
\end{align}
The dimensionless magnetic dipole moment, $\mu_m$, is obtained through the graph of $F_G(\theta)$ versus angle $\theta$, related by the formula $F_G(\theta)=\mu_m\sin\theta$. The dimensionless energy of the configuration is given by~\cite{r9}:
\begin{equation}
E = \frac{g}{8\pi\xi}  \int(B^a_i B^a_i + D_i\Phi^a D_i \Phi^a + \frac{\lambda}{2} (\Phi^a\Phi^a - \xi^2)^2)r^2\sin\theta ~drd\theta d\phi
\label{eq7}
\end{equation}

\section{The Magnetic Ansatz}
\label{sec:3}

The magnetic ansatz~\cite{r7} used in this paper in order to produce one-plus-half monopole is
\begin{eqnarray}
&& gA^a_i = - \frac{1}{r} \psi_1 \hat{n}^a_\phi\hat{\theta}_i + \frac{1}{r\sin\theta} P_1 \hat{n}^a_\theta\hat{\phi}_i + \frac{1}{r} R_1 \hat{n}^a_\phi\hat{r}_i - \frac{1}{r\sin\theta} P_2 \hat{n}^a_r\hat{\phi}_i, \nonumber \\
&& gA^a_0 = 0 \,,\,\ g\Phi^a = \Phi_1 \hat{n}^a_r + \Phi_2 \hat{n}^a_\theta,
\label{eq8}
\end{eqnarray}
where $P_1 = \sin\theta\psi_2$, $P_2=\sin\theta R_2$ and all the profile functions $\psi_1, \psi_2, R_1, R_2, \Phi_1$ and $\Phi_2 $ are functions of $r$ and $\theta$. The spatial spherical coordinate unit vectors are given by
\begin{eqnarray}
\hat{r}_i &=& \sin\theta ~\cos \phi ~\delta_{i1} + \sin\theta ~\sin \phi ~\delta_{i2} + \cos\theta~\delta_{i3}, \nonumber\\
\hat{\theta}_i &=& \cos\theta ~\cos \phi ~\delta_{i1} + \cos\theta ~\sin \phi ~\delta_{i2} - \sin\theta ~\delta_{i3}, \nonumber\\
\hat{\phi}_i &=& -\sin \phi ~\delta_{i1} + \cos \phi ~\delta_{i2}.
\label{eq.9}
\end{eqnarray}
and the isospin coordinate unit vectors are
\begin{eqnarray}
\hat{u}_r^a &=& \sin \theta ~\cos n\phi ~\delta_{1}^a + \sin \theta ~\sin n\phi ~\delta_{2}^a + \cos \theta~\delta_{3}^a,\nonumber\\
\hat{u}_\theta^a &=& \cos \theta ~\cos n\phi ~\delta_{1}^a + \cos \theta ~\sin n\phi ~\delta_{2}^a - \sin \theta ~\delta_{3}^a,\nonumber\\
\hat{u}_\phi^a &=& -\sin n\phi ~\delta_{1}^a + \cos n\phi ~\delta_{2}^a,
\label{eq.10}
\end{eqnarray}

The boundary conditons for $r$, when $r$ approaches infinity are:
\begin{eqnarray}\label{eq11}
&& \psi_1 = \frac{3}{2},~~ P_1=n\sin\theta+ \frac{n (1+\cos\theta)}{2}\sin \frac{\theta}{2} , \nonumber\\ 
&&  R_1 = 0, ~~P_2 = n\cos\theta- \frac{n (1+\cos\theta)}{2}\cos \frac{\theta}{2} , \nonumber\\
&& \Phi_1=\xi \cos \frac{\theta}{2} , ~~\Phi_2=\xi \sin \frac{\theta}{2}.
\end{eqnarray}
Near the origin, we have the common trivial vacuum solution:
\begin{eqnarray}\label{eq12}
&& \psi_1(0,\theta)=P_1(0,\theta)=R_1(0,\theta)=P_2(0,\theta)=0,\nonumber\\
&& \sin\theta~\Phi_1(0,\theta)+\cos\theta~\Phi_2(0,\theta)=0, \nonumber\\
&& \partial_r(\cos\theta\Phi_1(r,\theta)-\sin\theta\Phi_2(r,\theta))|_{r=0}=0.
\end{eqnarray}
The corresponding boundary conditons imposed along the positive and negative $z$-axis ($\theta = 0$ and $\theta = \pi$) are as follows:
\begin{eqnarray}\label{eq13}
\partial_\theta\Phi_1=\Phi_2=\partial_\theta\psi_1=R_1=P_1=\partial_\theta P_2(r,\theta)=0
\end{eqnarray}

Equations~\eqref{eq11}--\eqref{eq13} constitute the full set of boundary conditions for $r$ and $\theta$. Then, upon substituting the magnetic ansatz~\eqref{eq8} into the equations of motion~\eqref{eq3}, the set of equations of motion are reduced to six coupled second order partial differential equations. Those equations are solved numerically with the given boundary conditions using Maple and MATLAB~\cite{r11}. These six coupled second order partial differential equations were then transformed into a system of nonlinear equations using the finite difference approximation, and then discretized on a non-equidistant grid of size $110\times100$ covering the integration regions $0 \leq x \leq 1$ and $0 \leq \theta \leq \pi$. Here $x = r / (r + 1)$ is the finite interval compactified coordinate. Also, some constants, such as $g$ and $\xi$ were set to one in the process.

\section{Results and Discussion}
\label{sec:4}

\begin{figure}[tbh]
	\centering
	\hskip0in
	 \includegraphics[scale=0.56]{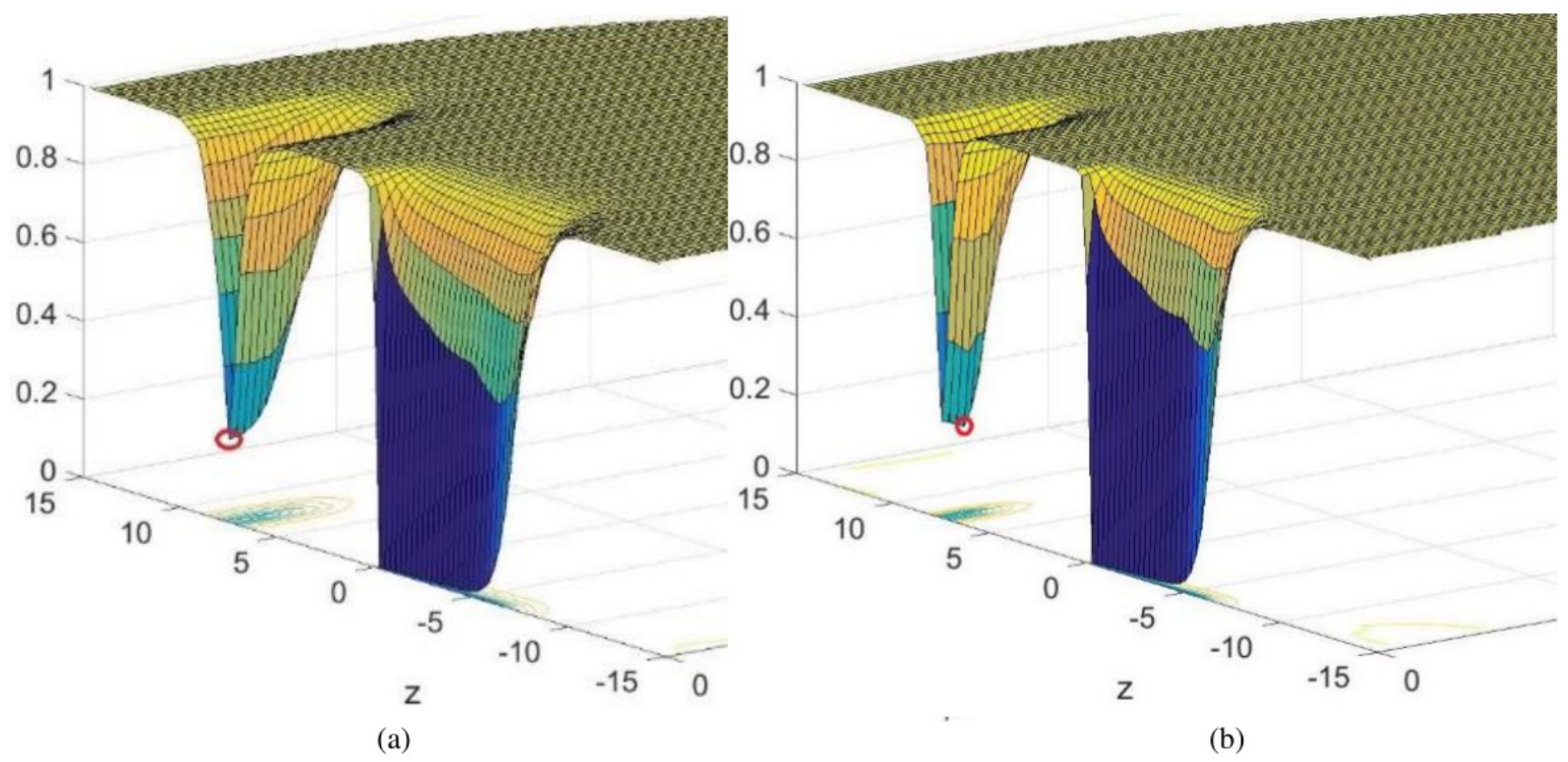} 
	\caption{\small 3D Higgs modulus plots for $n=3$ HEB solution when (a) $\lambda = 3$, (b) $\lambda = 10$. \label{figure1}  }
 \label{Fig.1}
\end{figure}

For $\phi$-winding number $n = 2$, $3$ and $4$, Higgs modulus, magnetic charge density and energy density are plotted. Physical quantites investigated in this research involve the separation between one-monopole and half-monopole, $d_z$, magnetic dipole moment, $\mu_m$, and total energy of the configuration, $E$.


\begin{footnotesize}
\begin{table}
\centering
\caption{Various critical values of Higgs coupling constants $\lambda$ corresponding to lower bound $\lambda_b$, bifurcation $\lambda_c$, and transition $\lambda_t$.}\label{table1}
\begin{tabular}{|c|c|c|c|}
\hline
    & Lower bound $\lambda_b$ & Bifurcation $\lambda_c$ & Transition $\lambda_t$ \\\hline
$n = 2$ FB & 1.96  &  - & -   \\\hline
$n = 3$ FB & 0.48  &  - & - \\\hline
$n = 3$ HEB & - & 2.28 & 3.32 \\\hline
$n = 3$ LEB & - & 2.28 & 8.83 \\\hline
$n = 4$ FB  & 0.57 & - & - \\\hline
$n = 4$ HEB & - & 2.87 & 3.62 \\\hline
$n = 4$ LEB & - & 2.87 & 9.12 \\\hline
$n = 4$ NB & 0.20 & - & - \\\hline
\end{tabular}
\end{table}
\end{footnotesize}


\begin{figure}[tbh]
	\centering
	\hskip0in
	 \includegraphics[scale=0.4]{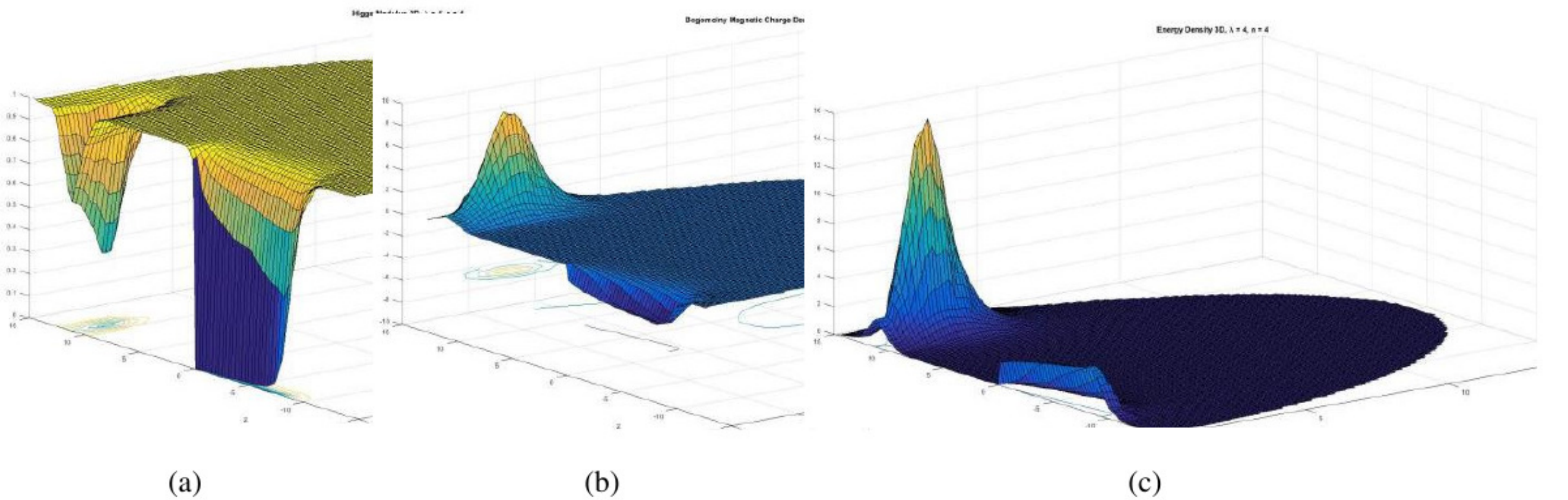} 
	\caption{\small Plots of the new branch (NB) soluttion of $n = 4, \lambda = 4$ for (a) Higgs modulus, (b) magnetic charge density and (c) energy density.\label{figure2} }
\end{figure}

The presence of a lower bound $\lambda_b$, in FB among all solutions, though, appears to be unexpected, they share similar characteristics with the critical points, $\lambda_c$, whereby below which no numerical solutions can be found. In fact, $\lambda_c$ themselves can be viewed as a type of lower bounds. Thus, like the critical points, the lower bound, $\lambda_b$, for all FB must arise due to the natural properties of the solution. Both $\lambda_b$ and $\lambda_c$ for all solutions are tabulated in Table~\ref{table1}.

In 3D Higgs modulus plots (Figure 1), half-monopoles are located at the origin, extend towards the negative $z$-axis and show a string-like formation. The 't Hooft-Polyakov monopoles are located on the positive $z$-axis as shown in Figure~\ref{figure1}(a). This is a standard multimonopole solution as the lowest point of Higgs modulus is located on the $z$-axis as indicated by the red circle. However, in Figure~\ref{figure1}(b), the lowest point is deviated from the axis and it is a clear indication that this is a vortex ring solution. Both Figure~\ref{figure1}(a) and Figure~\ref{figure1}(b) are plots for $\phi$-winding number $n = 3$ of HEB solution. In particular, Figure~\ref{figure1}(a) corresponds to $\lambda = 3$ while Figure~\ref{figure1}(b) corresponds to $\lambda = 10$. Clearly, a transition from multimonopole solutions to vortex ring solutions occurred somewhere between $\lambda = 3$ and $\lambda = 10$. The exact transition points are tabulated in Table~\ref{table1}. Similar transitions occurred in other branches and for different values of $n$ as well. A pattern can be obeserved from Table~\ref{table1}, transitions from multimonopole to vortex ring configuration occur only for HEB and LEB solutions for both $n = 3$ and $4$. Transitions occurred earlier in HEB solution relative to LEB solution. However, no transition is observed in $n = 4$ NB solution, this new branch is a vortex ring configuration for the entire branch. 

Plots of Higgs modulus, magnetic charge density and energy density for $n = 4$ NB solution are shown in Figure~\ref{figure2}. In Figure~\ref{figure2}(a), the position of the lowest point of the Higgs modulus is a clear indication of vortex ring structure. The magnetic charge of 't Hooft-Polyakov monopoles and half-monopoles are of opposite sign are shown in Figure~\ref{figure2}(b) with 't Hooft-Polyakov monopoles possess the positive charges. This feature is presented in all solutions. Plot of energy density is shown in Figure~\ref{figure2}(c).

\begin{figure}
\begin{center}
	\includegraphics[width=\linewidth,keepaspectratio]{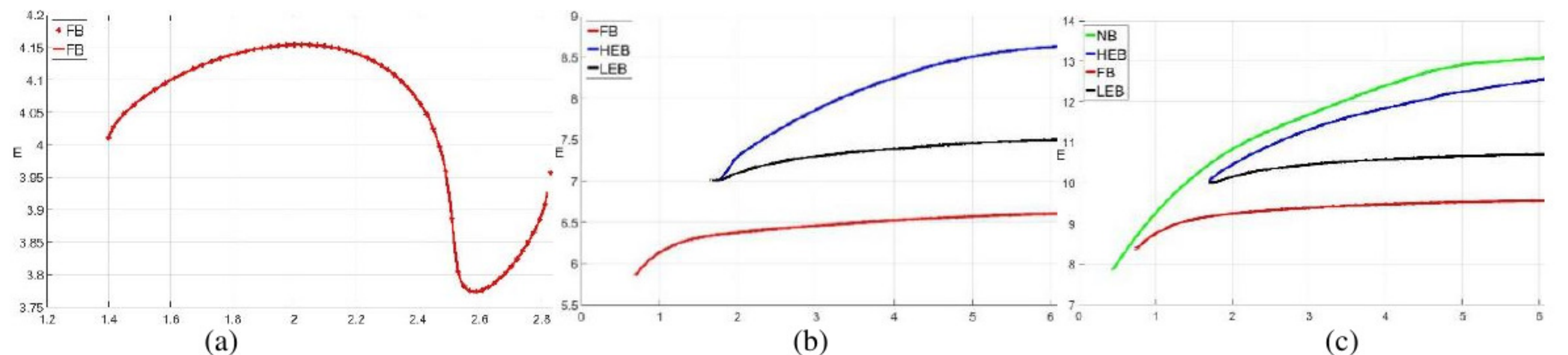}
\end{center}
\caption{\small The plots of total energy, $E$ versus $\lambda^{1/2}$, for (a) $n = 2$, (b) $n = 3$, (c) $n = 4$. \label{figure3}}
\end{figure}

Plots of total energy versus $\lambda^{1/2}$ is presented in Figure~\ref{figure3}. These plots show similar behaviours for $n = 3$ and $4$ with the exception that there is one more branch (NB) for $n = 4$ as can be seen in Figure~\ref{figure3}(b) and (c). For $ n = 2$, however, the shape of the curve in Figure~\ref{figure3}(a) is quite distinctive. It might indicate superimposing two half-monopoles would give rise to some unfamiliar physical processes. This certainlys require further investigation. In general, energy increases significantly with the change of $n$ and energy continues to increase as the Higgs coupling constant $\lambda$, increases. Figure 4 shows the corresponding magnetic dipole moment versus $\sqrt{\lambda}$ for the one-plus-half monopole for $n=2, 3$ and 4.

\begin{figure}
\begin{center}
	\includegraphics[width=\linewidth,keepaspectratio]{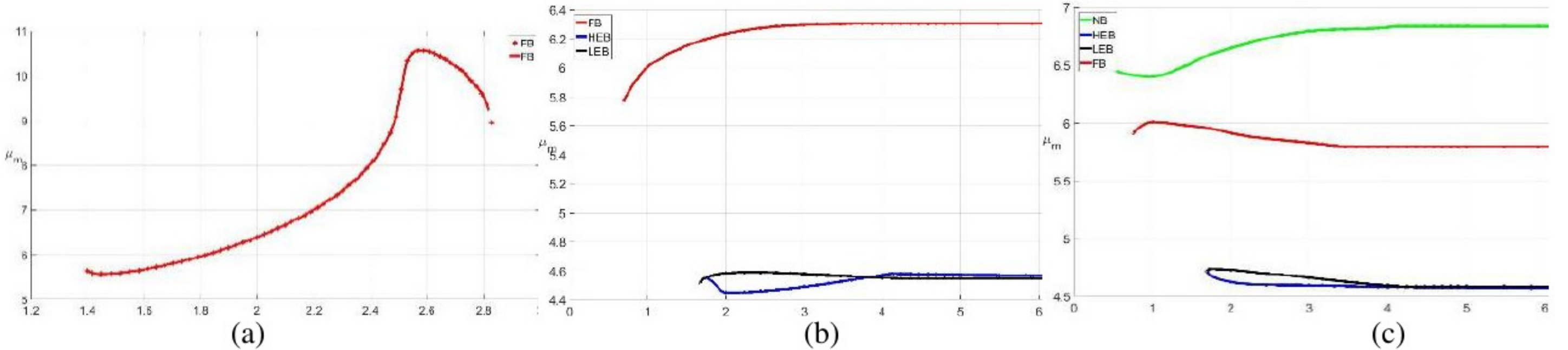}
\end{center}
\caption{\small The plots of magnetic dipole moment, $\mu_m$, versus $\lambda^{1/2}$, for (a) $n = 2$, (b) $n = 3$, (c) $n = 4$.  \label{figure4}}
\end{figure}

\section{Conclusions}
\label{sec:5}
As demonstrated above (Figure 3 and 4), there are bifurcating branches which emerged for both $n = 3$ and $4$ above the fundamental branch (FB, the red curve) with the blue curve being HEB solution and the black curve being LEB solution. For $n = 4$, however, a new branch appeared (green curve) with even higher energy. This is a vortex ring solution for the entire branch, no transitions occurred. Transitions from multimonopole solutions to vortex ring solutions occurred only for HEB and LEB of $n = 3$ and $4$, $\lambda_t$ are listed in Table~\ref{table1}. Lower bounds, $\lambda_b$ are presented as well. Besides a lower bound, there exists an upper bound as well for $n = 2$ as shown in Figure~\ref{figure3}(a), which requires further investigation. 

Recently MAP, MAC and vortex ring solutions has been constructed in SU(2) $\times$ U(1) Weinberg-Salam theory \cite{r12}. Hence our reuslts in this paper serves not only to to enrich our understanding in the SU(2) Yang-Mills-Higgs theory, but also serves as a stepping stone towards constructing such solution in a more realistic theory. These work will be reported in a future work. 

\section{Acknowledgement}
\label{sec:6}
The authors would like to thank School of Physics, Universiti Sains Malaysia.

\end{document}